\documentclass[twoside]{article}
\usepackage{fleqn,espcrc2}
\usepackage{feynmf}

\usepackage{graphicx}

\newcounter{saveeqn}
\newcommand{\alphaeqn}{\refstepcounter{equation}\setcounter{saveeqn}{\value{equation}}%
\setcounter{equation}{0}%
\renewcommand{\theequation}{%
        \mbox{\arabic{saveeqn}.\alph{equation}}}}%

\newcommand{\reseteqn}{\setcounter{equation}{\value{saveeqn}}%
\renewcommand{\theequation}{\arabic{equation}}}

\newcommand{\AmS}{{\protect\the\textfont2
  A\kern-.1667em\lower.5ex\hbox{M}\kern-.125emS}}

\newcommand{\hep}[1]{ [#1]}
 
\newcommand{\lsi}{\raise0.3ex\hbox{$<$\kern-0.75em\raise-1.1ex\hbox{$\sim$}}}
\newcommand{\gsi}{\raise0.3ex\hbox{$>$\kern-0.75em\raise-1.1ex\hbox{$\sim$}}}

\newcommand{\gvec}[1]{\mbox{\boldmath $#1$}}
\renewcommand{\vec}[1]{{\bf #1}}

\newcommand{\mlabel}[1]{\label{#1}}

\title{Non-equilibrium field theory\thanks{Supported by the TMR network 
``Finite temperature phase
transitions in particle physics'', EU contract no. ERBFMRXCT97-0122.} 
        }

\author{%
Dietrich B\"odeker\address[MCSD]{The Niels Bohr Institute,
        Blegdamsvej 17, DK--2100 Copenhagen
        }%
      \thanks{e-mail: bodeker@nbi.dk}
\hfill\raisebox{41mm}[0mm][0mm]{\makebox[0mm][r]{NBI-HE--00-45}}%
}
       
\begin{document}

\begin{abstract}
  I discuss various topics in relativistic non-equilibrium field
  theory related to high energy physics and cosmology. I focus on
  non-perturbative problems and how they can be treated on the
  lattice.  \vspace{1pc}
\end{abstract}

\maketitle
\section{INTRODUCTION}
\mlabel{sc:introduction}
The subject of non-equilibrium field theory is the evolution of
many-particle systems in real (Minkowski) time. It also gives new
insights into the much better understood equilibrium properties of
quantum fields. Most of its applications in high energy physics are
related to the early universe and to heavy ion collisions.

Many interesting problems in non-equilibrium field theory are 
non-perturbative. One could hope that they can be put on a lattice and
then be solved by a computer. However, lattice simulations of quantum
field theory (QFT) work in Euclidean time and nobody knows how to perform
real  time simulations.

A considerable simplification occurs for systems which are in partial
(or incomplete) equilibrium (see, e.g., \cite{Landau5}).  It means
that some slowly relaxing quantities $ y _ \alpha $ differ
significantly from their equilibrium values, while all other degrees
of freedom are thermalized. The most extreme example are hydrodynamic
modes, their relaxation times diverge when their wavelengths go to
infinity (other examples will be discussed in Sec.~\ref{sc:close}).
The non-equilibrium problem can then be reduced to evolution
equations for the $ y _ \alpha $. These equations contain so called
transport coefficients, and the task of non-equilibrium field theory
is to determine them from the underlying QFT.

Partial equilibrium is a good approximation for most of the history of
the early universe (an important exception is (p)reheating at the end
of inflation, that is the transition to a radiation dominated epoch,
see Sec.~\ref{sc:preheating}).  Needless to say that the most solid
results have been obtained for this case.

Fluctuation dissipation relations allow to express transport coefficients
in terms of thermal averages like  
\begin{eqnarray} 
        \langle [O(t), O(0)] \rangle 
        \mlabel{correlator}
        ,
\end{eqnarray}
where the angular brackets denote  the average over a thermal ensemble
with temperature $ T $,
\begin{eqnarray}  
  \langle \cdots \rangle \equiv Z ^{-1}  {\rm tr}(\cdots e ^{- H/T} )
  ;
\end{eqnarray}
$ H $ is the Hamiltonian.  For example, electric conductivity is
obtained from (\ref{correlator}) with $ O $ being the electromagnetic
current (color conductivity, on the other hand, has no interpretation
in terms of a gauge invariant correlation function of some
current. Instead it arises as a ``Wilson coefficient'' in an effective
theory for long distance modes of non-abelian gauge fields, see
Sec.~\ref{sc:effective}).

I have already mentioned that a serious difficulty of thermal field
theory is the fact that one has to deal with real  time.  Let
me illustrate this for the case of the expectation value
(\ref{correlator}).  One could think of expanding
Eq.~(\ref{correlator}) in powers of $ t $. Then one can compute every
coefficient in the expansion from the usual Euclidean path integral of
thermal QFT. But transport coefficients are determined by the small
frequency limit of the Fourier transform of (\ref{correlator}). This
depends on the behaviour for $ t \to \infty $   which can
not be captured by an expansion around $ t = 0 $.

Fortunately there is a limit of quantum field theory in which a
non-perturbative treatment is possible. It is the classical field
limit of bosonic fields. It applies when the number of quanta in the
field modes of interest is large.  This is indeed the case in a
variety of interesting problems. I discuss the classical field
approximation, together with some recent applications in
Sec.~\ref{sc:classical}. A wide class of approximations, many of them 
related to a large $ N $ expansion,  which are not
restricted to the classical field limit is discussed in
Sec.~\ref{sc:hartree}. In Sec.~\ref{sc:close} I consider systems in
incomplete equilibrium. I show how one can use the classical field
approximation even when quantum effects are important by constructing
effective classical field theories. Sec.~\ref{sc:summary} is a brief
summary of my talk.
\section{CLASSICAL FIELD APPROXIMATION}
\mlabel{sc:classical}
While quantum systems are impossible to simulate in real time, there
is no principle obstacle to performing real time simulations in
classical field theory.  All one has to do is to solve classical field
equations of motion with the appropriate initial conditions, and
determine the physical quantity of interest from the solution.

A well known special case of the classical field approximation
(usually not referred to as such) is the dimensional reduction of a $
d + 1 $ dimensional thermal QFT in the imaginary time formalism to the
$ d $ dimensional theory for the zero Matsubara frequency modes of the
bosonic fields \cite{dimreduction}. This is the formal classical limit
because for $ \hbar \to 0 $ the non-zero modes become infinitely heavy
and decouple. The thermodynamics of QFT can be studied with
4-dimensional lattice simulations, which is the only reliable way to
access the properties of hot QCD near the critical
temperature. Dimensional reduction is convenient for very high
temperatures when there is a large separation of the scale $ 2 \pi T $
and the screening length(s) of the system.  In non-equilibrium field
theory the role of the classical field approximation is much more
important since there is no tractable analogue of the 4-dimensional
Euclidean theory.

The classical field approximation should be reliable when the number
of field quanta in each relevant field mode is large. There are indeed
interesting applications where this is the case.  The remainder
of this section is a  list of some of them.

\subsection{Preheating after inflation}
\mlabel{sc:preheating}
Inflation is the only known solution to the horizon and flatness
problem in standard Big Bang cosmology. Furthermore, it generates
density fluctuations which can seed the structure formation once
electrons and nuclei have combined to form atoms. There has been a lot
of interest in the transition from an inflationary to the radiation
dominated epoch after it was realized \cite{Kofman} that the
initially homogeneous inflaton field $ \varphi $ can decay very rapidly
into low momentum modes of the inflaton itself or into modes of other
scalar fields $ \chi $ through parametric resonance. This mechanism is
referred to as pre-heating, and it has drastically changed the picture
of reheating after inflation.

The large amplitudes of the amplified modes make this problem non-perturbative.
At the same time, the occupation numbers of these modes grow very large
which opens the possibility to study this problem in the classical
field approximation on the lattice \cite{Son,Khlebnikov}. As an
illustration, Fig.~1 shows the occupation number of the $ \chi $ field
(taken from Ref.~\cite{Prokopec}). 

Lattice simulations have shown that non-linear effects play an
important role in reheating \cite{Khlebnikov,Prokopec}.  Effects like
non-thermal phase transitions and defect formation have been
investigated \cite{Kasuya:1997,Khlebnikov:nonthermal,Kasuya,Felder}.
Preheating and the possibility of non-thermal phase transitions in
gauge theories were studied in \cite{Rajantie}.
\begin{figure}
\includegraphics[width=18pc]{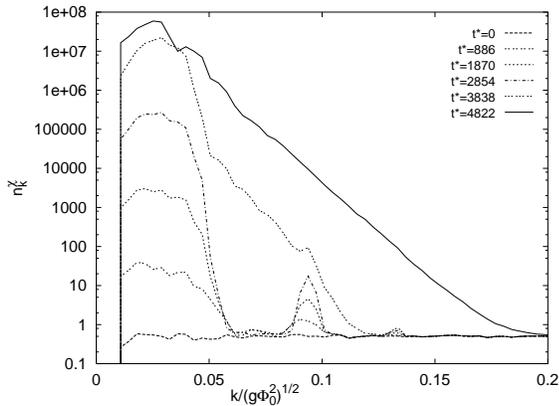}
\caption{Occupation numbers of $ \chi $-modes vs momentum.
The curves correspond to different values of time. There are
three visible resonances caused by the oscillating inflaton background 
(from Ref.~[6]).}
\end{figure}

\subsection{Small $ x $ gluon distributions}
The gluon density in hadrons and nuclei becomes large at small $ x $,
where $ x $ is the fraction of the momentum of the hadron/nucleus
which is carried by the gluon. Then the usual perturbative treatment
of parton evolution breaks down. In \cite{McLerran} a model was
proposed which describes the small $ x $ gluon field of large nuclei
classically.  It was first applied to nuclear collisions in
Ref.~\cite{Kovner}. If one is interested in the production of gluons
with small transverse momenta a non-perturbative treatment of this
model is necessary. A lattice version was developed in
\cite{Krasnitz:1999}.  It was used to solve the non-linear equations
of motion for the gluon fields of two colliding nuclei, and energy
densities and gluon multiplicities in the final state were extracted
\cite{Krasnitz}.

\subsection{Tests of approximation schemes}
\mlabel{sc:test}
The approximation schemes to be discussed in Sec.~\ref{sc:hartree} are
usually formulated in quantum field theory.  It is difficult to asses
their reliability. In Ref.~\cite{Aarts:thermalization} it was pointed
out that they can equally well be formulated in the classical field
limit.  It is then possible to perform lattice simulations and compare
the ``exact'' lattice results with the approximate ones. For a scalar
model in (1+1) dimensions it was found that early-time behaviour is
reproduced qualitatively by the Hartree approximation, and that
including scattering improves this to the quantitative level.  At late
times the lattice system thermalizes, but the improved Hartree
approximation fails to reproduce this behaviour \cite{Aarts:exact}.

\subsection{Droplet nucleation in a thermal first order phase transition}
Starting from a high temperature, a first order phase transition
proceeds through the nucleation of droplets, which then grow and
eventually fill the whole space with the low temperature phase.  The
droplet nucleation is caused by thermal fluctuations.  Most
calculations of the nucleation rate use the formalism due to Langer
\cite{Langer} which is based on a saddle point approximation around
the the critical droplet. A lot of lattice computations have been
performed, most of them have considered a moderately strong first
order phase transition (for recent results, see \cite{Borsanyi}).
Ref.~\cite{Moore:droplets} has considered a very strongly first order
transition in the electroweak theory.  The results were compared with
various analytic calculations based on Langer's approach. It was found
that the accuracy of the analytic results strongly depend on the
approximation used for the critical droplet action.  The best
agreement was found when the 2-loop effective potential for the Higgs
field \cite{Arnold;Espinosa,Fodor;Hebecker} together with the field
dependent wave function renormalization \cite{Bodeker:Z,Schmidt} is
used.

\section{BEYOND THE CLASSICAL FIELD APPROXIMATION}
\mlabel{sc:hartree}
The classical field approximation is clearly restrictive, and it is
not capable of treating genuine quantum effects. For example, the
thermalization of the universe after preheating (cf.\
Sec.~\ref{sc:close}) is not classical. It is sometimes possible to
obtain effective classical theories when the non-classical modes are
weakly interacting and can be integrated out in perturbation theory
(see  Sec.~\ref{sc:effective}). In general, however, other
approximations are necessary.

The Hartree approximation was discussed in the talk by J.~Vink
\cite{Vink:talk} and there are several related approaches. Usually the basic
idea is to work directly with non-equilibrium Green functions or,
alternatively, with the generating functional of 1-PI Green functions
\cite{Wetterich}.  These satisfy certain Schwinger-Dyson-like
equations which are an infinite set of equations containing the full
set of $ n $-point functions.  To make this problem tractable, the
hierarchy of Schwinger-Dyson equations must be truncated.  It is
possible to formulate such truncations as a systematic expansion in $
1/N $, where $ N $ is the number of some fields
\cite{largeN,Boyanovsky}.

Recent lattice studies have focused on the question whether one can
obtain thermalization staring from some non-equilibrium initial state
(see also Sec.~\ref{sc:test}). Thermalization can be obtained in the
Hartree approximation if the mean field is allowed to be
inhomogeneous, which was discussed in J.~Vink's talk \cite{Vink:talk}.
Thermalization of fermionic modes, which can
never be described by classical fields, was also observed in the
inhomogeneous mean field approximation \cite{Aarts:inhomogeneous}.

Ref.~\cite{Berges} considers  (1+1) dimensional $ \lambda \varphi ^ 4 $ theory 
in the loop expansion of the generating functional of
2-particle irreducible non-equilibrium Green functions. At 3-loop
order, when scattering is taken into account, thermalization was
observed.

\section{CLOSE TO THERMAL EQUILIBRIUM}
\mlabel{sc:close}
As I mentioned in the introduction, there are interesting
non-equilibrium processes in which most of the degrees of freedom are
equilibrated and only some slowly relaxing ones are not.  The task of
non-equilibrium field theory is then to determine the relevant
diffusion coefficients.

In weakly coupled theories it is possible to compute some transport
coefficients using perturbation theory. This is not an easy task.
Typically one has to sum an infinite set of diagrams, which was done
for the shear and bulk viscosity in (3+1) dimensional scalar theory
with cubic and quartic self interactions \cite{Jeon}. The result of
\cite{Jeon} is equivalent to the solution of the classical Boltzmann
equation with appropriate coefficients. Other transport coefficients
were calculated starting directly from the Boltzmann equation, which
is technically much simpler than a diagrammatic analysis
\cite{Arnold:transport}.  Another application of the Boltzmann
equation  of current interest is the generation of a lepton
asymmetry due to heavy Majorana neutrinos \cite{Buchmuller}.

Unfortunately, it is not clear whether
the classical Boltzmann equation is the leading order effective theory
in some systematic weak coupling expansion which would allow to go
beyond the leading order.  There are problems where this is necessary,
like for example in electroweak baryogenesis \cite{Rubakov}.

There are transport coefficients which are not computable in
(appropriately resummed) perturbation theory even when the coupling
constant is small. A famous example is the rate $ \gamma _ {B + L} $
for baryon ($ B $) plus lepton number ($ L $) dissipation in the standard
electroweak theory \cite{Rubakov}. The Lagrangian of the electroweak
theory has a U(1) symmetry which would imply that $ B + L $ is
conserved. However, this symmetry is anomalous, and $ B + L $
violating processes are possible.  Suppose there is some non-zero
density $ N _ {B+L} $ in the universe. It then gets washed out
according to
\begin{eqnarray}
  \frac{ d }{d t} N _ {B+L} = - \gamma _ {B+ L} N _ {B+L}
  .
\end{eqnarray} 

There is a fluctuation dissipation relation stating that $ \gamma _
{B + L} $ is proportional to the Chern-Simons diffusion rate $ \Gamma
_ {\rm CS} $ of the weak SU(2) gauge fields.  $ \Gamma _ {\rm CS} $ is
dominated by soft gauge field modes with typical momenta of order $ g
^ 2 T $. This is the so called magnetic screening scale at which
perturbation theory for hot Yang-Mills theory breaks down
\cite{Linde}. Fortunately the smallness of this scale allows the use
of the classical field approximation to compute $ \Gamma _ {\rm CS} $.
The occupation number of the bosonic field modes can be estimated
from the Bose-Einstein distribution function
\begin{eqnarray}
  n(\vec k ) = \frac{ 1}{e ^{|\vec k|/T} -1} \simeq \frac{ T}{|\vec k|}
    \quad (|\vec k| \ll T)
\end{eqnarray}
which is of order $ 1/g ^ 2 $ for the soft ($|\vec k| \sim  g
^ 2 T $) modes.

The question how to calculate $ \Gamma _ {\rm CS} $ has a long history.
One reason for this long lasting interest is that the $ B + L $
dissipation rate is a fundamental quantity of the Standard Model of
electroweak interactions (SM) and extensions thereof. It plays an
important role in particle physics scenarios for generating the baryon
asymmetry of the universe.  The use of classical lattice field theory
for calculating $ \Gamma _ {\rm CS} $ was first suggested in
Ref.~\cite{Grigoriev}, where a (1+1) dimensional model was studied.

Originally it was assumed that the naive classical field approximation
correctly describes the dynamics of the soft modes in (3+1)
dimensional Yang-Mills theory. Therefore the lattice version
of the classical Yang-Mills equations of motion
\alphaeqn
\mlabel{classicalYM}
\begin{eqnarray}
  D _ 0 \vec E + \vec D \times \vec B & = & 0 
\mlabel{classicalYM.a}
        ,
\\
  \vec D \cdot \vec E &=& 0
\mlabel{classicalYM.b}
        ,
\end{eqnarray}
\reseteqn
were solved and used to determine the Chern-Simons diffusion
\cite{Ambjorn,Ambjorn;Krasnitz,Tang} (see also \cite{mr}). Here $ \vec
E $ and $ \vec B $ are the color \footnote{I use ``color'' as a
generic term for some non-abelian gauge charge, not as something
specific to QCD.} electric and magnetic fields and $ D _\mu $ denotes
the covariant derivative $ \partial _ \mu + g A _ \mu $ in the adjoint
representation. The above assumption would be correct if high momentum
($ |\vec k| \gg g ^ 2 T $) modes decouple from the soft dynamics. It is
suggestive because it is true for the thermodynamics of hot Yang-Mills
fields: At leading order in $ g $ the soft non-perturbative modes are
described by the 3-dimensional pure Yang--Mills theory of dimensional
reduction, and the only role of high momentum modes is to make the
coupling constant equal to its renormalized value.

\subsection{Problems of the classical field approximation}
\mlabel{sc:problems}
Classical thermal field theory as a continuum theory does not
exist. \footnote{Take for example the energy density of a free
massless bosonic field at temperature $ T $. It is proportional to $
T \Lambda ^ 3 $, where $ \Lambda $ is the UV cutoff.}  This has been
known for more than 100 years and it has played a crucial role in the
discovery of quantum theory by Planck.

In dimensional reduction the UV divergences are taken care of by
adding the appropriate local counterterms to the 3-dimensional action.
Then long distance ($ l \gg T ^{-1} $) correlation functions of the
4-dimensional theory are correctly reproduced by the 3-dimensional one.

However, it was realized in \cite{bms} that there are UV divergences
in classical thermal Yang-Mills theory which do not occur in equal
time correlation functions \footnote{Real time correlation functions
in scalar $ \lambda \varphi ^ 4 $ theory, on the other hand, are
renormalized by the counterterms of dimensional reduction
\cite{Aarts:finite}.}. These divergences are closely related to what
is called ``hard thermal loops'' in thermal QFT \cite{htl}.  They are
particularly ugly, they are non-local in space and time and on the
lattice they are sensitive to the lattice geometry (see also
\cite{Arnold:lattice}).  The occurrence of such divergences in lattice
field theory is possible due to the lack of Lorentz invariance  which
is explicitly broken by the presence of the thermal bath. As a
consequence, Green functions with external momenta $ p _ \alpha $
depend not only on the Lorentz invariants $ p _ \alpha \cdot p _ \beta
$ but on frequencies and spatial momenta separately.

\begin{figure}
\vspace{1cm}
\begin{fmffile}{gluonloop}
\begin{fmfchar*}(60,60)
  \fmfleft{links} 
  \fmf{photon,tension=5}{links,lv}\fmf{photon,tension=5}{rv,rechts}
  \fmf{photon,left}{lv,rv,lv}  
  \fmfright{rechts}
  \fmfdot{lv,rv}
\end{fmfchar*}
$\quad$
\begin{fmfchar*}(60,60)
    \fmfleft{links} 
    \fmf{photon,tension=5}{links,mv}\fmf{photon,tension=5}{mv,rechts}
    \fmf{photon,right=90,tension=.5}{mv,mv}  
    \fmfright{rechts}    
  \fmfdot{mv}
\end{fmfchar*}
$\quad$
\begin{fmfchar*}(60,60)
    \fmfleft{links} 
    \fmf{photon,tension=5}{links,lv}\fmf{photon,tension=5}{rv,rechts}
    \fmf{dbl_dots,left}{lv,rv,lv}  
    \fmfright{rechts}    
  \fmfdot{lv,rv}
\end{fmfchar*}
\end{fmffile}

\caption{One-loop polarization tensor in non-abelian gauge theory. Only the
  first two diagrams contribute to the UV divergence in the classical
  field limit, while the ghost loop contribution is suppressed.}
\end{figure}
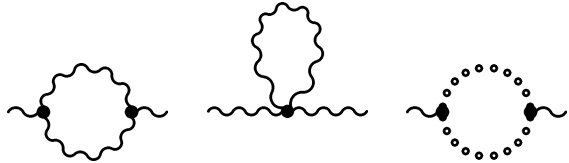

Since such divergences are probably not familiar to most of the
audience, let me describe them in some more detail.  They are due to
the one-loop polarization tensor (see Fig.~2).  First consider the
case that the external momentum $ k $ is purely spatial, $ k _ 0 = 0
$. This is what one encounters in the 3-dimensional reduced theory.
The result is of the form
\begin{eqnarray}
  \Pi^{\mu \nu } (0,\vec k)  
  \sim  g ^ 2 T \int ^ \Lambda  \frac{d ^ 3 q }{\vec q ^ 2} 
  f \left( \frac{ \vec k \cdot \vec q}{\vec q ^ 2}, 
  \frac{ \vec k ^ 2}{\vec q ^ 2 } \right) 
  \mlabel{pi.3d}
\end{eqnarray}
where the function $ f $ is dimensionless, and $ \Lambda $ is the UV
cutoff. $ f $ can be expanded in powers of its arguments, and only the
lowest order term $ f (0,0) $ gives rise to a linearly divergent
integral,
\begin{eqnarray}
  \Pi^{\mu \nu } (0, \vec k)  \sim  g ^ 2 T  \Lambda f(0,0) 
  + \mbox{UV finite}
  \mlabel{div1}
  .
\end{eqnarray}
Gauge invariance ensures that this divergence occurs only in $ \Pi ^
{00} $.  This is nothing but the familiar divergence of the mass term
for the adjoint Higgs field $ A _ 0 $ which is taken care of by adding
a mass counterterm to the 3-dimensional action.

Now consider the analogous contribution when the external frequency is
non-zero. In the classical field limit the result for $ \Pi^{\mu \nu
}$ is similar to Eq.~(\ref{pi.3d}).  However, the function $ f $ now
depends on three dimensionless variables,
\begin{eqnarray}
  \Pi^{\mu \nu } (k) 
  \sim  g ^ 2 T \int ^ \Lambda  \frac{d ^ 3 q }{\vec q ^ 2} 
  f \left( \frac{ k ^ 0}{\hat {\vec q} \cdot \vec k}, 
    \frac{ \vec k \cdot \vec q}{\vec q ^ 2}, 
  \frac{ \vec k ^ 2}{\vec q ^ 2 }\right) 
        .
        \mlabel{pi.classical}
\end{eqnarray}
To extract the divergent contribution one can again expand in powers of
$ \vec k \cdot \vec q / \vec q ^ 2$  and $ \vec k ^ 2 / \vec q ^ 2 $, 
but the dependence on the first argument of $ f $ remains. Therefore
the linearly divergent part of $ \Pi^{\mu \nu } $ has a non-trivial dependence
on $  k ^ 0/\hat {\vec q} \cdot \vec k $, 
\begin{eqnarray}
  \Pi  ^{\mu \nu }(k)\sim  g ^ 2 T \Lambda 
  \int _{\hat {\vec q}} f \left( \frac{ k ^ 0}{\hat {\vec q} \cdot \vec k}, 
    0,0 \right) 
  + \mbox{UV finite}
  \mlabel{div2}
  .
\end{eqnarray}
(I've been somewhat sketchy, for the precise form of the divergent
contribution on the lattice see \cite{bms,Arnold:lattice}.)  Such a
divergence is clearly a disaster because it is non-local in
space and time.\footnote{In QFT theory the linear divergences in
  Eqs.~(\ref{div1}), (\ref{div2}) are cut off by the Bose-Einstein
  distribution function.} Effects of these divergences in lattice
simulations have first been observed in
Refs.~\cite{Bodeker:plasmon,Tang:plasmon}.

It was argued in \cite{asy} that the UV divergent part of $ \Pi ^{\mu
\nu }$ has the same qualitative effect as the hard modes in the
quantum theory (see Sec.~\ref{sc:effective}), which is that it slows
down the dynamics of the soft gauge field modes. In the continuum
limit $ \Lambda \to \infty $ the characteristic time scale of the soft
modes should diverge, so that the Chern-Simons diffusion rate $ \Gamma
_ {\rm CS} $ should go to zero.  This behaviour was indeed observed in
a recent calculation on a very fine lattice \cite{mr}.

\subsection{Effective classical theories}
\mlabel{sc:effective}
If the high momentum modes, which do not have large occupation numbers
and are thus not classical, are so important for the soft field
dynamics, can one still make use of the classical field approximation?
This is indeed possible. The modes with momenta larger than $ g ^ 2 T
$ are weakly coupled and can be integrated out using perturbation
theory.  One finds that the correct effective classical theory, at
leading order in $ g $ and $ \log(1/g) $, is described
by the equation of motion \cite{langevin} 
\begin{eqnarray}
  \vec D \times \vec B = \gamma \vec E + \gvec{ \zeta }
  \mlabel{langevin}
\end{eqnarray}
instead of Eq.~(\ref{classicalYM}). The constant $ \gamma $ is
proportional to $ T/\log(1/g) $. Furthermore, $ \gvec{ \zeta } $ is a
Gaussian white noise\footnote{``White'' refers to the frequency
  spectrum of $ \gvec{ \zeta }$ being flat.}. Its expectation value
vanishes and it is entirely determined by its 2-point function
\begin{eqnarray}
  \langle \zeta _ i ^ a (x)\zeta _ j ^ b (x')\rangle 
  = 2 T \gamma \delta ^{ab}
  \delta _ {ij} \delta ^{(4)}(x - x')
  .
\end{eqnarray}
Here $ a $, $ b $ are indices of the adjoint representation of the
gauge group. Recently the leading log effective theory
(\ref{langevin}) was extended to include the Higgs field
\cite{Moore:higgs}.

The physics behind Eq.~(\ref{langevin}) is easily understood. The soft
gauge fields evolve in time, which means that the color electric field
is non-zero.  The hot plasma is a conductor and the electric field
induces a current which is carried by charged particles corresponding
to the hard ($ \vec k \sim T $) field modes. The particles suffer
collisions with a typical mean free path of order $ (\log(1/g) g ^ 2
T)^{-1} $. In the weak coupling limit $ \log(1/g) \gg 1 $ this is
small compared to the wavelengths of the soft non-perturbative
modes. Thus the soft modes cannot resolve the trajectories of the
particles and the induced current is simply a number times the
electric field.  The noise term is due to thermal fluctuations of the
field modes which have been integrated out.

Eq.~(\ref{langevin}) is similar to the equation of motion of an
ordinary (abelian) magnetic field in a conducting plasma, where the
role of $ \gamma $ is played by the electric conductivity. (In analogy
with electrodynamics one calls the coefficient in Eq.~(\ref{langevin})
color conductivity). The key difference is that the electrodynamics
analogue of Eq.~(\ref{langevin}) is valid to arbitrary accuracy if one
considers fields with longer and longer wavelengths. This is possible
because abelian magnetic fields are not screened. On the other hand,
non-abelian magnetic fields are screened on a length scale of order $
(g ^ 2 T ) ^{-1} $. Thus there is nothing like a magneto-hydrodynamic
limit in hot non-abelian plasmas. In fact, Eq.~(\ref{langevin})
is only valid at leading and, as was shown
recently in Ref.~\cite{ay:nll}, at next-to-leading order in $
\log(1/g) ^{-1} $.

From Eq.~(\ref{langevin}) one can estimate the characteristic
frequency $ k _ 0 $ of the soft modes as follows. The rhs is of order
$ (g ^ 2 T )^ 2 \vec A $, since it contains two covariant derivatives
of the vector potential $ \vec A $. The electric field which contains
one time derivative can be estimated as $ k _ 0 \vec A $. 
Thus the characteristic frequency is
\begin{eqnarray}
  k _ 0 \sim \log(1/g) g ^ 4 T
  \mlabel{frequency}
  .
\end{eqnarray}
This is much smaller than the estimate $ k _ 0 \sim g ^ 2 T $ which
one would obtain assuming that high momentum modes decouple from the soft
dynamics.
From Eq.~(\ref{frequency}) one can estimate the Chern-Simons diffusion
rate, which is the number of topological transitions per unit time
and unit volume,  as
\begin{eqnarray}
  \Gamma _ {\rm CS} \sim k _ 0 |\vec k| ^ 3 \sim \log(1/g) \alpha ^ 5 T ^ 4
  \mlabel{estimate}
	,
\end{eqnarray}
where $ \alpha = g ^ 2 /(4 \pi) $.

Eq.~(\ref{frequency}) was obtained through a sequence of effective
field theories by integrating out more and more high momentum degrees
of freedom. The first is the so called 

{\em Hard thermal loop effective theory}\\
It is the result of integrating out ``hard'' physics associated with $
k ^ 2 $ of order $ T ^ 2 $, and it can be described by the classical
field equations of motion \cite{Blaizot:93} 
\alphaeqn
 \mlabel{htl}
\begin{eqnarray}
          D _ 0 \vec E + \vec D \times \vec B & = & m^2_{\rm D} \vec W  
        \mlabel{htl:ampere}
\\
  \vec D \cdot \vec E &=& m^2_{\rm D} W ^ 0
\mlabel{htl:gauss}
\\
         v \cdot D  W &=& \vec{v}\cdot\vec{E} 
        \mlabel{htl:c} 
        .
\end{eqnarray}  
\reseteqn 
The field $ W = W(x,\vec v) $ represents the fluctuations of adjoint
color charge due to the hard degrees of freedom, which act like
particles with 3-velocity $ \vec v $, $\vec v ^ 2 =1 $.
Furthermore,
\begin{eqnarray} 
 v ^ \mu = (1, \vec v)
 .
\end{eqnarray}
The 4-current on the rhs of Eqs.~(\ref{htl:ampere}) and
(\ref{htl:gauss}) is given by
\begin{eqnarray}
  W ^ \mu (x)= \int \frac{ d ^ 2 \Omega _ {\vec v}}{4 \pi}
   v ^ \mu W(x,\vec v)
  ,
\end{eqnarray}
times the square of the leading order Debye mass $ m_{\rm D} \propto g
T $.  Originally hard thermal loops where discovered when trying to
develop a consistent perturbative expansion of Greens functions with
external momenta of order $ g T $ \cite{htl}.  The effect of hard
thermal loops on the soft dynamics was first realized in
Ref.~\cite{asy} where the fields in Eq.~(\ref{htl}) where interpreted
as being soft. Then the main effect of the hard particles is Landau
damping \cite{Landau10} of the soft modes, resulting in the estimate $
k _ 0 \sim g ^ 4 T $, which is the same as Eq.~(\ref{frequency}) but
without the logarithm.  It was then realized that the fields in
Eq.~(\ref{htl}) must be interpreted as modes with both $ \vec k \sim g
T $ and $ \vec k \sim g ^ 2 T $ \cite{langevin}.

{\em Non-abelian Boltzmann equation}\\ 
The $ \vec k \sim g T $ modes in Eq.~(\ref{htl}) are weakly coupled
and can be integrated out in an expansion in $ g $.  They are
responsible for small angle scattering between the hard particles.
They induce a collision term and noise term in Eq.~(\ref{htl:c}), and
one obtains \cite{langevin}
\alphaeqn%
\mlabel{maxwellboltzmann}
\begin{eqnarray}
          D _ 0 \vec E + \vec D \times \vec B & = & m^2_{\rm D} \vec W  
        \mlabel{boltzmann:ampere}
\\
  \vec D \cdot \vec E &=& m^2_{\rm D} W ^ 0
\mlabel{boltzmann:gauss}
\\
         ( C + v \cdot D)  W &=& \vec{v}\cdot\vec{E} + \xi
        \mlabel{boltzmann} 
        ,
\end{eqnarray}  
\reseteqn
where now the gauge and $ W $ fields contain only spatial Fourier
components of order $ g ^ 2 T $ (modulo logarithms of $ 1/g $). The
linear collision term $ C W $ breaks time reflection invariance and
describes dissipation caused by the modes which have
been integrated out.  These modes also perform thermal fluctuations
which pump energy back into the $ g ^ 2 T $ modes through the Gaussian
white noise $ \xi $.  (For other approaches, which do not make use of
the hard thermal loop effective theory, see
Refs.~\cite{asy2,ValleBasagoiti,Litim,Blaizot:boltzmann}.)

{\em Langevin equation(s)}\\ 
The collision term $ C $ in
Eq.~(\ref{boltzmann}) is of order $ \log(1/g) g ^ 2 T $, while
derivatives of the non-perturbative modes are of order $ g ^ 2 T $. At
leading order in $ \log(1/g) ^{-1} $ the latter can be neglected. Then
the current on the rhs of Eq.~(\ref{boltzmann:ampere}) is a local and
linear function of $ \vec E $ and $ \xi $. Due to the strong damping
one can neglect the kinetic term $ D _ 0 \vec E $ in
Eq.~(\ref{boltzmann:ampere}) which leads to
Eq.~(\ref{langevin}). 

Starting from a non-local Langevin equation
which is valid to all order in $ \log(1/g)^{-1} $
\cite{Arnold:langevin} it was shown that the logarithmic approximation
can be systematically improved in an expansion in $ \log(1/g)^{-1} $
\cite{ay:nll}.  It was found that Eq.~(\ref{langevin}) is still valid
at next-to-leading order in $ \log(1/g)^{-1} $ if one includes a
next-to-leading log correction in the color conductivity $ \gamma $,
which was computed in \cite{ay:nll}.
\begin{figure}
\includegraphics[width=16pc]{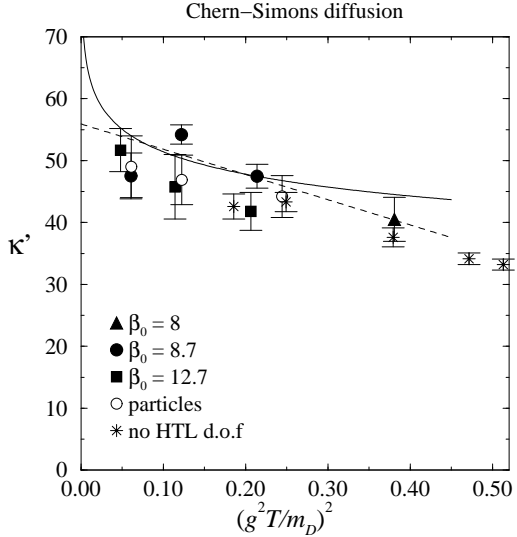}
\caption{Chern-Simons diffusion rate parametrised by $ \kappa ' $ (see
Eq.~(\ref{rate})) vs the magnetic screening scale over Debye mass (for
details see text).
}
\end{figure}

\subsection{Lattice results}
\mlabel{sc:lattice}
The effective theory (\ref{langevin}) is perfectly suited for lattice
simulations because it is UV finite \cite{asy2}. It was used to
calculate the Chern-Simons diffusion rate in
Ref.~\cite{Moore:log}. Due to Eq.~(\ref{estimate}) the result can be
parametrized as
\begin{eqnarray}
  \Gamma _{\rm CS} 
  = \kappa'
  \frac{ g ^ 2 T ^ 2}{m^2_{\rm D}} \alpha ^ 5 T ^ 4 
  \mlabel{rate}
\end{eqnarray}
with
\begin{eqnarray}
  \kappa' = \kappa  \log \left( \frac{ m^2_{\rm D} }{g ^ 2 T} \right)
  \mlabel{grappa}
  .
\end{eqnarray} 
The dimensionless constant $ \kappa  $ was determined as
\begin{eqnarray}
  \kappa  = 10.8 \pm 0.7
  .
\end{eqnarray}
Near a weakly first order electroweak phase transition one has to take
into account the Higgs field which requires an extension of
Eq.~(\ref{langevin}) \cite{Moore:higgs}, the resulting  effect 
on $ \Gamma _{\rm CS} $ is  of order 20\%.

The hard thermal loop effective theory (\ref{htl}) is a classical
field theory which correctly incorporates the effect of the hard
modes. Unlike the Langevin equation (\ref{langevin}) its validity is
not restricted to the logarithmic approximation, but it does not have
a continuum limit. \footnote{In the UV the effect of the $ W $ fields
can be ignored, and Eq.~(\ref{htl}) has the same UV problems as
classical thermal Yang-Mills theory.} It was argued that it should still
be  useful for lattice simulations as long as the lattice spacing
is not too small \cite{particles}.

Two different lattice implementations of the hard thermal loop
effective theory have been developed. In Ref.~\cite{Hu} the hard modes
are represented by classical charged particles.  This method was
further developed and used to compute $ \Gamma _{\rm CS} $ in
Ref.~\cite{particles}. Starting point of Ref.~\cite{bmr} were the
equations of motion (\ref{htl}). The $ W $ fields were expanded in
spherical harmonics and only components $ W _ {lm} $ with $ l $
smaller or equal to some cutoff $ l _ {\rm max} $ were kept.

The results are shown in Fig.~3. The rate $\Gamma _{\rm CS} $ is
parametrized by $ \kappa' $ through Eq.~(\ref{rate}).  The
black symbols are the results of Ref.~\cite{bmr} for 3 different
values of the lattice spacing $ a $, where
\begin{eqnarray}
  \beta _ 0 \equiv \frac{ 4}{g ^ 2 T a}
  .
\end{eqnarray}
The open circles are the results of Ref.~\cite{particles}. Also shown
are the results of Ref.~\cite{mr} for the classical Yang-Mills theory
(\ref{classicalYM}). \footnote{For the results of Ref.~\cite{mr} the
  value of $ m^2_{\rm D} $ in Eqs.~(\ref{rate}) and (\ref{grappa}) was
  chosen such that the classical polarization tensor (\ref{div2}),
  averaged over the directions of $ \vec k $, has the same $ k _ 0 \to 0$
  asymptotics as the physical hard thermal loops, following a
  suggestion in Ref.~\cite{Arnold:lattice}.} The different methods
give surprisingly similar results.  The dashed line is a linear fit $
\kappa ' = c _ 1 + c _ 2 \left( \frac{ g ^ 2 T}{m_{\rm D}} \right)^ 2
$. The full line is a fit which includes the leading log term, $ \kappa '
= \kappa \log \left( \frac{ m^2_{\rm D} }{g ^ 2 T} \right) + c$.

The results shown in Fig.~3 are about a factor 5 larger than the
leading log result. However, including the next-to-leading log
correction to the color conductivity $ \gamma $ increases the value of
$ \kappa ' $ in Eq.~(\ref{grappa}) by about a factor 5 \cite{ay:nll},
which gives  a remarkable agreement with the results in  Fig.~3.

\section{SUMMARY}
\mlabel{sc:summary}
At present we see an interesting interplay between analytic and
numerical lattice work on non-equilibrium field theory. It is
motivated by the physics of the early universe and of heavy ion
collisions.  Real time simulations of non-equilibrium quantum field
theory are not possible. A common tool is the classical field
approximation which can be applied if the field modes of interest have
large occupation numbers. There are several approaches which go beyond
the classical field approximation. They require other approximations,
such as Hartree-Fock, large $ N $.  It is possible to use the
classical field approximation even when quantum effects are important
by constructing effective classical theories.  These can differ
substantially from the naive classical limit of the underlying quantum
field theory.

\end{document}